# Strong Thermal Transport Anisotropy and Strain Modulation in Single-Layer Phosphorene

*Zhun-Yong Ong*[#(a)], *Yongqing Cai*[#], *Gang Zhang*[(b)], *and Yong-Wei Zhang*

Institute of High Performance Computing, A*STAR, Singapore 138632

[(a)] E-mail: ongzy@ihpc.a-star.edu.sg; [(b)] E-mail: zhangg@ihpc.a-star.edu.sg

**ABSTRACT**: Using first-principles calculations and non-equilibrium Green's function method, we investigate the ballistic thermal transport in single-layer phosphorene. A significant crystallographic orientation dependence of thermal conductance is observed, with room temperature thermal conductance along zigzag direction being 40% higher than that along armchair direction. Furthermore, we find that the thermal conductance anisotropy with the orientation can be tuned by applying strain. In particular, the zigzag-oriented thermal conductance is enhanced when a zigzag-oriented strain is applied but decreases when an armchair-oriented strain is applied; whereas the armchair-oriented thermal conductance always decreases when either a zigzag- or an armchair-oriented strain is applied. The present work suggests that the remarkable thermal transport anisotropy and its strain-modulated effect in single-layer phosphorene may be used for thermal management in phosphorene-based electronics and optoelectronic devices.

# INTRODUCTION

Two-dimensional (2D) layered crystals in their single-layer form have attracted considerable attention because of the unique physical properties associated with their reduced dimensionality, which can potentially be exploited for applications in nanoscale electronic and photonic devices. For example, graphene has by far the highest carrier mobility due to its massless charge carriers[1], while single-layer $MoS_2$, a member of the transition metal dichalcogenides (TMDs) family, has been regarded as a promising candidate for field effect transistor applications given its high on/off ratio[2].

Very recently, black phosphorus (BP)[3], a layered material consisting of sheets of $sp^3$-hybridized phosphorus atoms puckered along the so-called in-plane armchair direction and held together by weak van der Waals forces, has been the focus of a considerable amount of research into its electronic[4], optical[5], transport[6] and structural properties[7]. Part of this growing interest in BP may be attributed to the successful isolation of the few-layer forms of BP for spectroscopic and electrical characterization[6b] as well as to the much larger direct band gap demonstrated in few-layer BP[6]. Like other ultrathin two-dimensional (2D) crystals[8], few-layer BP also allows for superior electrostatic modulation of the carrier density, a feature that is necessary for continued nanoscale transistor scaling. Already, preliminary angle-resolved transport studies of few-layer BP show a high and significantly anisotropic hole mobility[6a, 9] as well as large and anisotropic in-plane optical conductivity at room temperature[5]. The single-layer form of BP, sometimes called phosphorene, is predicted to be a superior candidate material for field-effect applications given its atomic thinness and relatively high carrier mobility especially in the armchair direction[6a].

Thermal management in nanoscale electronic and optoelectronic devices remains a crucial issue,[10] especially as strong localized Joule heating in the confined volume of an ultrathin channel can reduce device reliability and performance. The use of a high thermal conductivity material in or around the transistor can help dissipate waste heat more efficiently, which is necessary for preventing device performance degradation and breakdown. Thus, the integration of a new material into a nanoscale transistor requires taking into account its thermal conduction properties. To this end, heat conduction in graphene[11] and in $MoS_2$[12] has been extensively studied. However, compared to these more widely studied 2D materials, the thermal transport properties of monolayer phosphorene are still not well-understood.

In this work, using the non-equilibrium Green's function (NEGF) method combined with first-principles calculations, we investigate ballistic thermal transport in monolayer phosphorene. In stark contrast to the isotropic thermal conductance in 2D graphene and $MoS_2$ sheets, thermal transport in monolayer phosphorene exhibits a strong orientation dependence. We also study how the thermal conductance varies with the magnitude and direction of the applied tensile strain. Our present study not only reveals physical insight into the thermal transport in monolayer phosphorene and but also provides promising guideline for thermal management in phosphorene-based nanoscale devices.

**RESULTS AND DISCUSSION**

**Phonon Dispersion**

Structurally, bulk BP is a layered material similar to graphite but with each layer consisting of an undulating array of ridges parallel to the *zigzag* direction, resembling an accordion[13]. Hence, its mechanical properties are highly anisotropic. The elastic stiffness constants and sound velocities have been measured to be much larger along the zigzag direction than along the armchair direction in bulk BP.[13-14] The linear compressibility is also much higher in the armchair direction than in the zigzag direction[15]. Likewise, we expect the phonon properties of single-layer phosphorene to be similarly orientation-dependent.

Figure 1 shows the phonon dispersion in the Γ-X (zigzag) and Γ-Y (armchair) directions. Around the Γ-point, the longitudinal acoustic (LA) phonon velocity is significantly higher in the zigzag direction because of the greater stiffness in the zigzag direction, as suggested by the accordion-like structure. This difference in the phonon velocities implies that there would be significant orientation dependence in the thermal transport, with the thermal conductance in the zigzag direction being greater.

**Anisotropic Thermal Conductance under Biaxial Strain**

Figure 2a shows the thermal conductance in the zigzag ($G_{zigzag}$) and armchair ($G_{armchair}$) direction at different biaxial tensile strain values (0, 2 and 4 percent) as a function of temperature from $T$=10 to 1000 K. At room temperature ($T$=300 K), $G_{zigzag}$ and $G_{armchair}$ for phosphorene with 0 percent strain equal to 1.08 and 0.76 nW/K/nm$^2$, respectively. The value for $G_{zigzag}$ is comparable to that for MoS$_2$ (~1.28 nW/K/nm$^2$)[16] but much lower than that for graphene (~4.1 nW/K/nm$^2$)[17]. This low thermal conductance may be due to the non-planar structure of the single-layer phosphorene. More interestingly, in stark contrast to the isotropic thermal conductance observed in monolayer graphene and MoS$_2$ sheets[12a-c], monolayer phosphorene exhibits a significant anisotropy in its thermal conductance. Figure 2b shows the thermal conductance ratio $G_{zigzag}/G_{armchair}$ as a function of temperature. At room temperature, $G_{zigzag}$ is about 40 percent larger than $G_{armchair}$, *i.e.* $G_{zigzag}/G_{armchair} \approx 1.4$. The anisotropy ratio increases as the temperature decreases, attaining a peak value of 1.6 at around 28 K at zero strain.

This anisotropy in thermal conduction has implications for heat dissipation in 2D crystal-based nanoscale devices where Joule heating in the confined volume of the atomically thin crystal can create localized hot spots[18] and the management of waste heat is expected to be a design bottleneck to the large-scale transistor integration into functional circuit. In graphene which has a high thermal conductivity (~2000 to 5000 W/K/m), Joule heat is expected to diffuse towards the metal electrodes, which act as heat sinks[18a, c]. On the other hand, in monolayer phosphorene, the carrier mobility is considerably higher along the armchair direction than along the zigzag direction. Thus in a phosphorene field effect transistor, the source-drain electrical current should be aligned in the armchair direction, which is unfortunately also the direction with the inferior thermal conductance, around 20% of that of graphene. Thus, new design strategies for the thermal management of phosphorene-based electrical and optoelectronic devices are needed. The relatively higher thermal conductance along the zigzag direction suggests that heat dissipation should be channeled transversely rather than to metal source/drain electrodes. The cross-shape device structure as shown in Fig. 2c allows the electrical current to flow along armchair direction while enabling heat to be dissipated transversely, taking advantage of the superior thermal conductance in the zigzag direction.

Next, we explore the effect of biaxial strain on the thermal conductance of monolayer phosphorene. Figure 2a shows both $G_{zigzag}$ and $G_{armchair}$ decreasing as a function of strain (0 to 4 percent). The thermal conductance reduction with increasing tensile strain is consistent with that observed in graphene[19] and other nanostructures[20]. However, the decrease is proportionally larger for $G_{armchair}$ than for $G_{zigzag}$. In the zigzag direction, the change is much smaller with the conductance change between 2 and 4 percent strain being almost imperceptible. Figure 2b shows the thermal conductance ratio $G_{zigzag}/G_{armchair}$ as a function of temperature. At 300 K, the $G_{zigzag}/G_{armchair} \approx 1.4$ for zero strain. However, as the strain increases to 4 percent, the ratio rises to 1.6 because $G_{armchair}$ decreases by a larger relative extent. This indicates that the conductance anisotropy can be enhanced by applying a biaxial strain. We also observe that at low temperatures ($T < 100$ K), the anisotropy peaks at around 28 K with a value of 1.6 for zero strain, rising to 2.0 at 4 percent strain.

**Anomalous Effects under Uniaxial Strain**

The change in thermal conductance under biaxial strain shows that the strain affects $G_{armchair}$ more than $G_{zigzag}$. To understand the mechanism more deeply, we apply a uniaxial strain in each direction and observe the change in thermal conductance. Figure 3 shows the thermal conductance in the zigzag ($G_{zigzag}$) and armchair ($G_{armchair}$) direction as a function of temperature ($T$=10 to 1000 K) for strain applied in the **(a)** zigzag and **(b)** armchair direction. $G_{armchair}$ decreases as we increase the strain in either the armchair or the zigzag direction. As expected, the decrease for uniaxial strain is not as great as the decrease for biaxial strain where we apply strain along both directions simultaneously. This also shows that the change in $G_{armchair}$ is relatively insensitive to the direction of the applied strain. The decrease in $G_{armchair}$ under tensile strain is similar to the trend observed in thermal transport simulations of bulk silicon and diamond, which has been explained as originating from the lower phonon group velocities when strain is applied[20].

On the other hand, we find that $G_{zigzag}$ exhibits an anomalous orientation-dependent change. When the strain is applied in the armchair direction, the $G_{zigzag}$ decreases like $G_{armchair}$. However, when the tensile strain is applied in the zigzag direction, $G_{zigzag}$ actually *increases* by a small but perceptible amount. This anomaly in the change in $G_{zigzag}$ and $G_{armchair}$ implies that the thermal conductance anisotropy can be modulated by changing the direction of the applied strain. Figure 4 shows the anisotropy ratio ($G_{zigzag}/G_{armchair}$) as a function of the applied strain (0 to 5 percent) in the zigzag and armchair direction. The anisotropy ratio is ~1.4 for zero strain, but as we increase the strain, the ratio rises. This rise in the anisotropy ratio is substantially greater for strain in the zigzag direction than in the armchair direction.

**Transmittance Analysis**

To understand the nontrivial strain effects on thermal conductance, we plot the phonon transmittance along the zigzag direction ($\Xi_{zigzag}(\omega)$, Fig. 5a) and armchair direction ($\Xi_{armchair}(\omega)$, Fig. 5b), respectively, for strain applied in the zigzag direction. In the plot of the transmittances, the spectrum gives the frequency-dependent relative contribution to the thermal conductance. We find that the strain in the zigzag direction results in markedly phonon softening which we observe from the red-shift in the high-energy ($\omega>350$ cm$^{-1}$) optical phonon transmittance. Another consequence of the strain is that the low frequency ($\omega<120$ cm$^{-1}$) $\Xi_{zigzag}$ is enhanced. In contrast, when the strain is applied in the armchair direction, the frequency redshift in $\Xi_{zigzag}(\omega)$ [Fig. 5c] and $\Xi_{armchair}(\omega)$ [Fig. 5d] are much smaller. There is also a small but visible reduction in the phonon transmittances, which explains the decrease in $G_{zigzag}$ and $G_{armchair}$ in Fig. 3b.

To understand how strain affects the relative contribution from phonons at different frequencies, we plot in Fig. 6 the differential thermal conductance, which is defined as:

$$\Delta G(\omega) = \frac{\hbar\omega}{2\pi S}\frac{df}{dT}\Delta\Xi(\omega),$$

where ΔΞ is the difference in the transmittance at 2 and 0 percent strain, for thermal transport in the **(a)** zigzag and **(b)** armchair direction. We set the temperature to be 300 K. $\Delta G(\omega)$ measures the *contribution* to the change in total thermal conductance by phonon modes with frequency ω, with respect to the 2 percent tensile strain (zigzag or armchair). A large $\Delta G(\omega)$ corresponds to a significant contribution to the change in thermal conductance by phonons with frequency ω.

It is interesting to find that with uniaxial strain along either the armchair or zigzag direction, the contribution by the high-frequency (ω>350 cm$^{-1}$) optical phonons to $\Delta G_{zigzag}$ and $\Delta G_{armchair}$ is negligible. With strain in the armchair direction, most of the change in the thermal conductance ($\Delta G_{zigzag}$ and $\Delta G_{armchair}$) is due to the drop in the low-frequency acoustic phonon contribution (60 to 260 cm$^{-1}$ for $\Delta G_{zigzag}$ and 60 to 140 cm$^{-1}$ for $\Delta G_{armchair}$). Thus, the reduced acoustic phonon transmittance is responsible for the decrease in the thermal conductance (both $G_{armchair}$ and $G_{zigzag}$) with strain in the armchair direction. On the other hand, when tensile strain is applied in the zigzag direction, there is an increase in the acoustic and optical phonon contribution between 0 and 260 cm$^{-1}$ to $G_{zigzag}$. The opposite trend is observed for $G_{armchair}$: there is a corresponding drop in the acoustic and optical phonon contribution over the same frequency range. Thus, with strain along zigzag direction, $G_{zigzag}$ increases with strain while $G_{armchair}$ changes only slightly.

**Summary**

We have found a strong anisotropy in thermal conductance of single-layer phosphorene: thermal conductance is much larger in the zigzag direction than in the armchair direction. Under the application of tensile strain in the armchair direction, the thermal conductance in the armchair and zigzag direction ($G_{armchair}$ and $G_{zigzag}$) decreases as expected. However, when the strain is in the zigzag direction, the $G_{armchair}$ decreases while $G_{zigzag}$ increases. This anomalous strain dependence in $G_{zigzag}$ can be attributed to the increase in enhanced contribution of the low-frequency phonon modes when the strain is applied in the zigzag direction. Combined the reported high carrier mobility along the armchair direction and the observed high thermal conductance along the zigzag direction in the present work, a cross-shape device structure is suggested for the thermal management of phosphorene-based electrical and optoelectronic devices. Moreover, our results also suggest that thermal transport in single-layer phosphorene can be modulated via applying tensile strain.

**METHODS**

In undoped single-layer phosphorene, thermal transport is largely mediated by phonons because its relatively low electrical conductivity[6b] implies that electrons only have a minor contribution to thermal transport. In a phonon-dominated system, the quantum thermal conductance can be calculated using the standard non-equilibrium Green's function (NEGF) method[21]. According to the well-known Landauer theory, the phonon-derived thermal conductance can be expressed as[21]:

$$G(T) = \frac{1}{2\pi S} \int_0^\infty d\omega\, \hbar\omega \frac{df}{dT} \Xi(\omega), \qquad (1)$$

where $S$ is the cross-sectional area with the strain in the transverse direction being taken into account and $f$ is the Bose-Einstein distribution function. The thickness of the phosphorene layer is assumed to be 0.55 nm. The transmittance or transmission function $\Xi(\omega)$ is calculated using the method described in Ref. [22] which sums over the transmittance contribution by the tranverse Fourier modes.

In this work, the interatomic force constants for all the phosphorous structures are calculated from first-principles calculation based on density functional theory (DFT) by using the QUANTUM ESPRESSO package[23]. We employ the local density functional together with the Trouiller-Martins-type[24] norm-conserving pseudopotentials. A cutoff energy of 50 Ry for the plane waves and a 14×10×1 Monkhorst-Pack $k$-point are adopted. Each phosphorous slab is separated by a vacuum of 15 Å normal to the surface. The atomic coordinates and the lattice constants are relaxed until the forces exerted on the atoms are less than 0.01 eV/Å and the stress less than 0.01 Kbar. At zero strain, the lattice constants are $a_1$ = 3.263 Å and $a_2$ = 4.341 Å. The dynamical matrix based on a 8×6×1 grid of $q$ points is calculated by using the density functional perturbation theory (DFPT), and the real space force constants are then obtained by the inverse Fourier transform of the dynamical matrix at each $q$ grid.

AUTHOR INFORMATION

#Both authors contributed equally to the work.

ACKNOWLEDGMENT

We acknowledge the funding support from the Agency for Science, Technology and Research, Singapore.

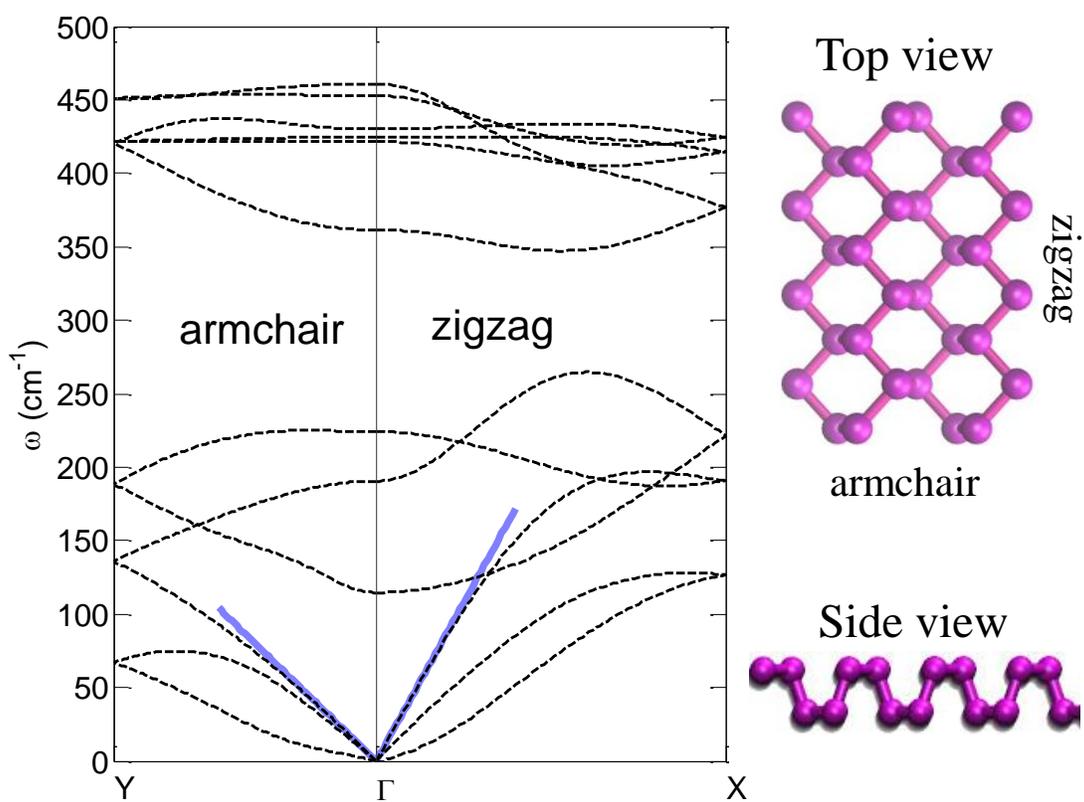

**Figure 1:** Phonon dispersion along the Γ-X (zigzag) and Γ-Y (armchair) direction in single-layer phosphorene. The longitudinal acoustic phonon velocity is higher in the zigzag direction. The side view shows the accordion-like structure along the armchair direction

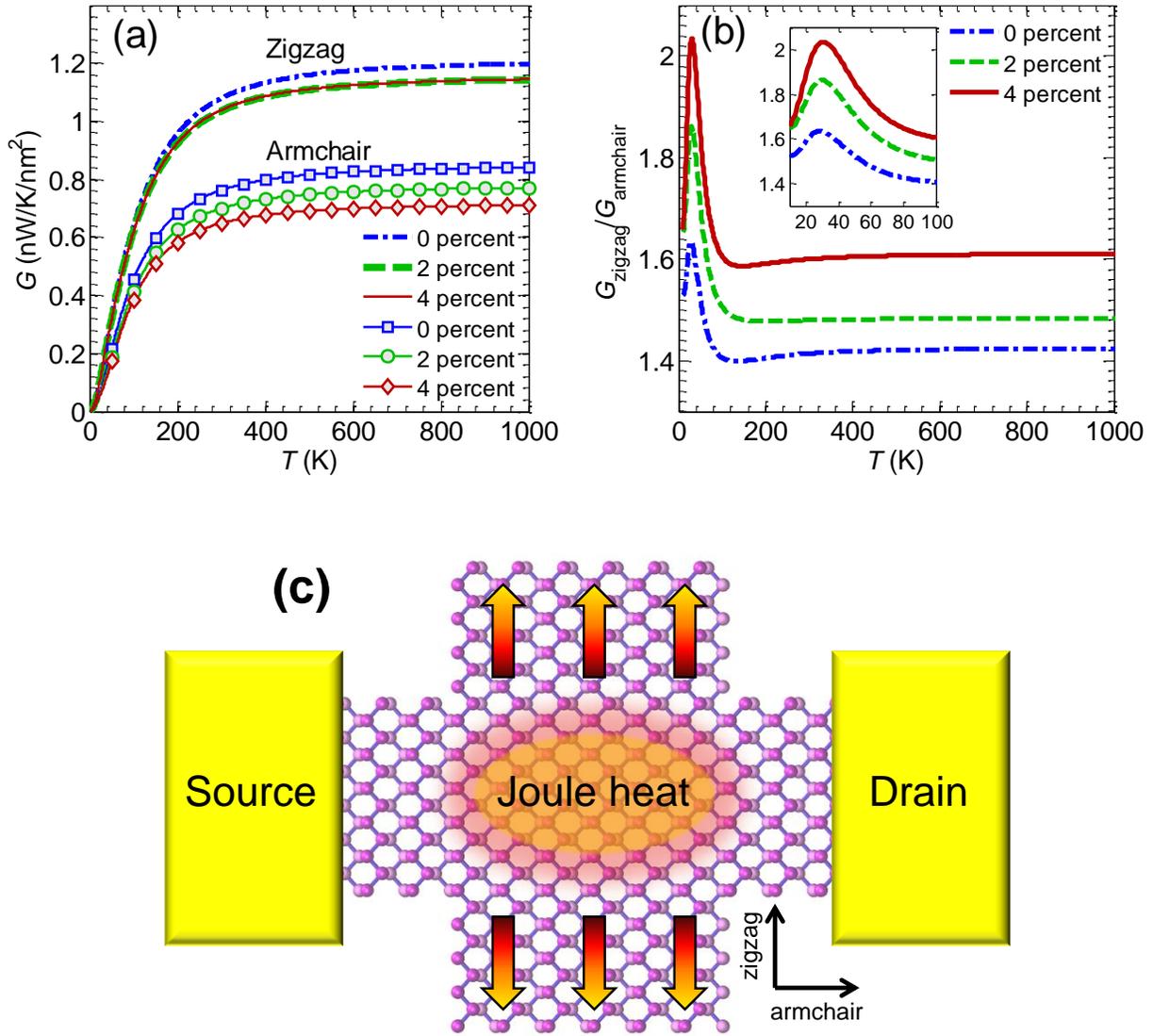

**Figure 2:** **(a)** Plot of the thermal conductance in the zigzag ($G_{zigzag}$) and armchair ($G_{armchair}$) direction as a function of temperature at different strain values (0, 2 and 4 percent) for biaxial strain. **(b)** Plot of their ratio ($G_{zigzag}/G_{armchair}$) as a function of temperature. The inset shows the ratio at low temperature. **(c)** Schematic of the cross-shape device structure for dissipating heat away in the direction transverse to the electrical current.

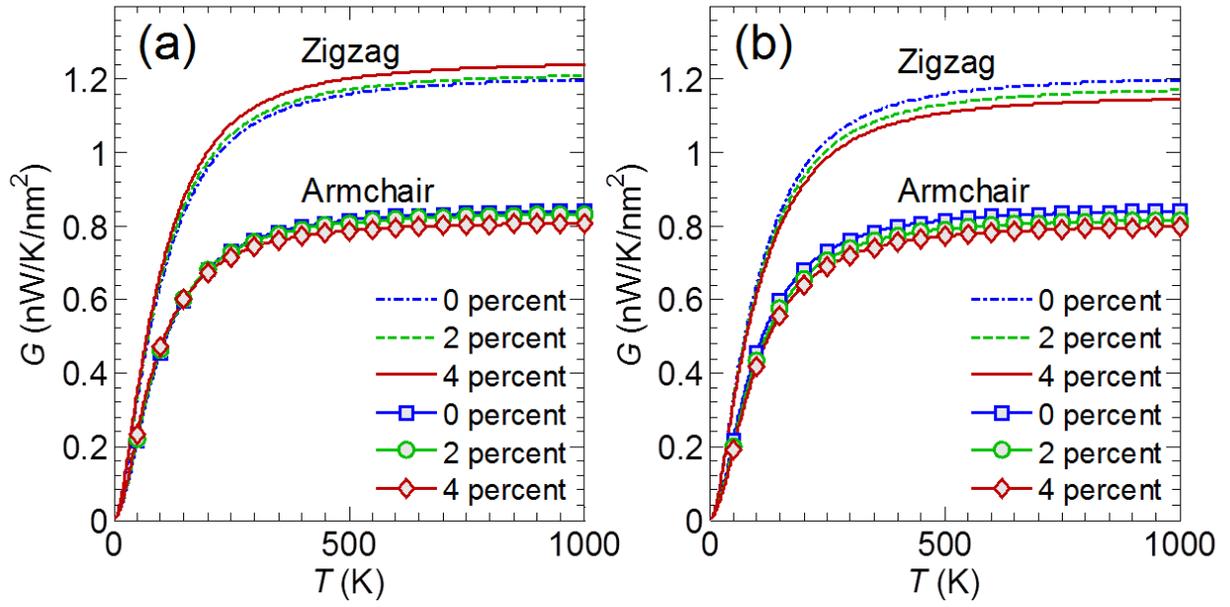

**Figure 3:** Plot of the thermal conductance in the zigzag ($G_{\text{zigzag}}$) and armchair ($G_{\text{armchair}}$) direction as a function of temperature at different strain values (0, 2 and 4 percent) for uniaxial strain applied in the **(a)** zigzag and **(b)** armchair direction.

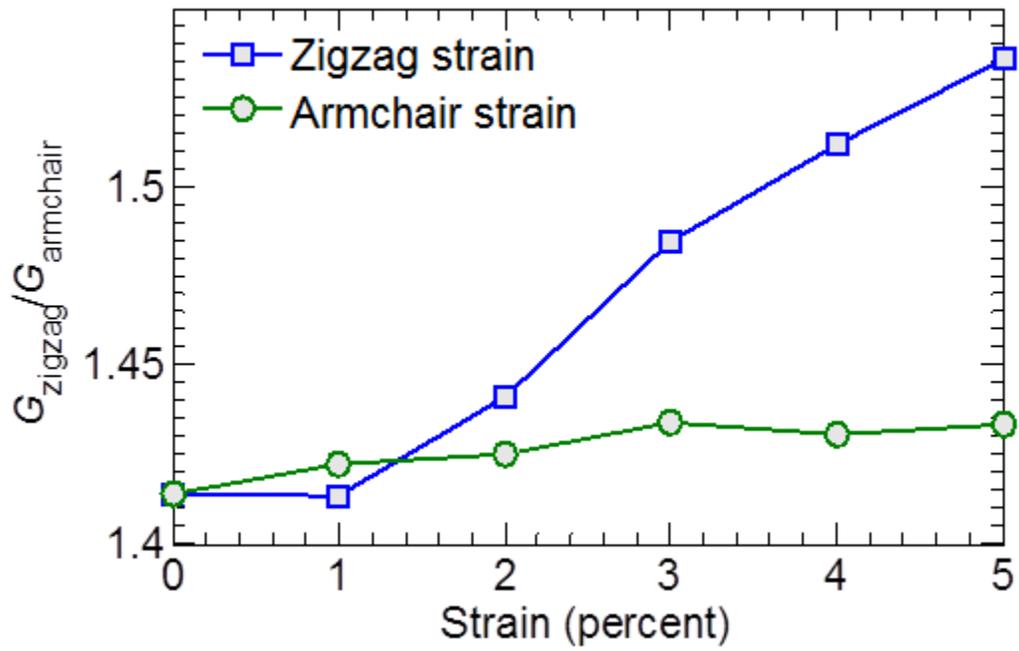

**Figure 4:** Plot of the thermal conductance anisotropy ($G_{zigzag}/G_{armchair}$) at different strain values (0 to 5 percent) for uniaxial strain applied in the zigzag (square) and armchair (circle) direction at 300 K.

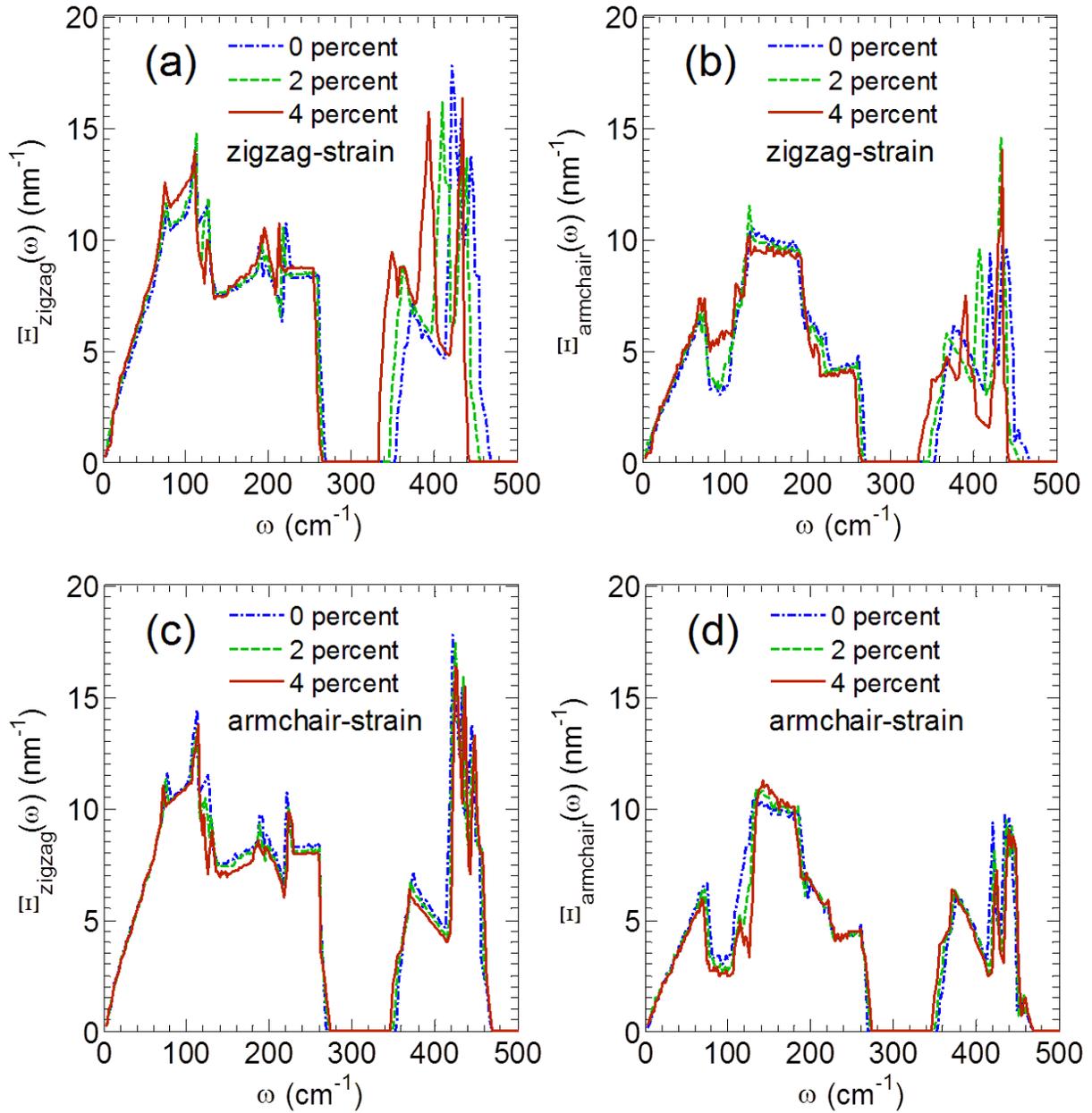

**Figure 5:** Plot of the zigzag and armchair-oriented cross-sectional phonon transmission ($\Xi_{zigzag}$ and $\Xi_{armchair}$) as a function of frequency at different strain values (0, 2 and 4 percent) for uniaxial strain applied in the **(a, b)** zigzag and **(c, d)** armchair direction.

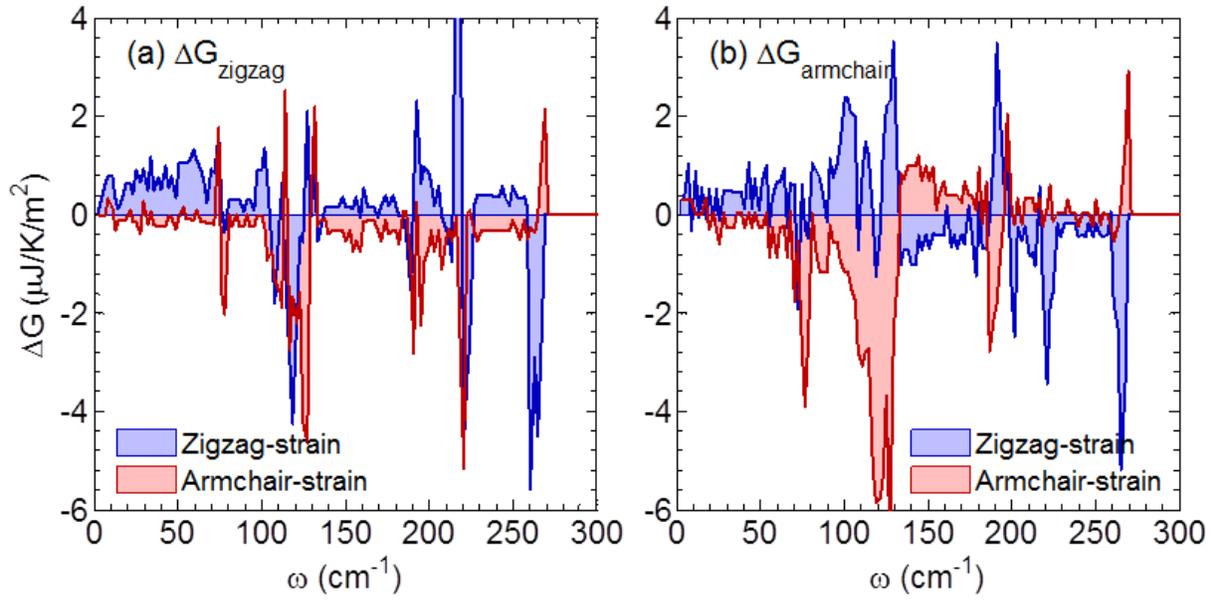

**Figure 6:** Plot of the differential thermal conductance ΔG for thermal transport in the **(a)** zigzag and **(b)** armchair direction, under the tensile strain in the zigzag and armchair directions.